# Enhancing the ultrafast third order nonlinear optical response by charge transfer in VSe$_2$-reduced graphene oxide hybrid

Vinod Kumar[1], Afreen[1], K. A. Sree Raj[2], Pratap mane[3], Brahmananda Chakraborty[4,5]

Chandra S. Rout[2], and K. V. Adarsh[1*]

[1]Department of Physics, Indian Institute of Science Education and Research, Bhopal 462066
India

[2]Centre for Nano and Material Science, Jain University Jain global campus, Jakkasandra,
Ramanagaram, Banglore 562112 India

[3]Seismology Division, Bhabha Atomic Research Centre, Trombay, Mumbai-400085, India

[4]High Pressure and Synchrotron Radiation Physics Division, Bhabha Atomic Research
Centre, Trombay, Mumbai 400085, India

[5]Homi Bhabha National Institute, Mumbai 400085, India

Nonlinear optical phenomena play a critical role in understanding microscopic light-matter interactions and have far-reaching applications across various fields, such as biosensing, quantum information, optical switching, and all-optical data processing. Most of these applications require materials with high third-order absorptive and refractive optical nonlinearities. However, most materials show weak nonlinear optical responses due to their perturbative nature and often need to be improved for practical applications. Here, we demonstrate that the charge donor-acceptor hybrid of VSe$_2$-reduced graphene oxide (rGO) hybrid exhibits enhanced ultrafast third-order absorptive and refractive nonlinearities compared to the pristine systems, at least by one order of magnitude. Through density functional theory and Bader charge analysis, we elucidate the strong electronic coupling in the VSe$_2$-rGO hybrid, involving the transfer of electrons from VSe$_2$ to rGO. Steady-state and time-resolved photoluminescence (PL) measurements confirm the electronic coupling and charge transfer. Furthermore, we fabricate an ultrafast optical limiter device with better performance parameters, such as an onset threshold of 2.5 mJ cm$^{-2}$ and differential transmittance of 0.42.

## INTRODUCTION

Third-order nonlinear optical effects play a crucial role in understanding the light-matter interaction, and is characterized by the susceptibility ($\chi$) = n$_2$ + i$\beta$ in which the real part (n$_2$) representing the nonlinear refractive index and the imaginary part ($\beta$) indicating the two-photon/excited state absorption[1–3]. These third-order nonlinear optical phenomena encompass significant effects such as third harmonic generation[4], intensity-dependent refractive index[5], optical Stark effect[6], and self-focusing or defocusing[7,8]. These phenomena hold great promise in various applications, including quantum information processing where they can serve as ultrafast optical switches[9], facilitating the generation of entangled photon pairs[10], optical modulators and limiters[11,12], saturable absorbers[13], and two-photon imaging and microscopy[14]. As a result, third-order optical nonlinearity has been extensively investigated in various materials, ranging from metals[15], semimetals[16], semiconductors from low dimension (0-2) to bulk[17–19], organic materials[20] and topological materials[21]. Examples of materials with good third-order nonlinear optical response include: graphene oxide (GO)[22], single- and multi-wall carbon nanotube[23,24], transition metal dichalcogenides (TMDs)[25], amorphous chalcogenide glasses[26], plasmonic materials[27] and quantum dots[17]. Unfortunately, conventional materials typically exhibit weak third-order nonlinear optical responses, even under intense illumination owing to its perturbative nature, while a strong response is crucial for applications[8]. For instance, materials exhibiting large nonlinear absorption coefficients and refractive index change at low intensities are highly desirable for photonics and optoelectronics applications. This emphasizes the need to develop new nonlinear optical materials with strong absorptive and refractive third-order nonlinearity. One capable technique to accomplish large third-order nonlinearities on a is through the photoinduced charge transfer of a donor-acceptor pair complex. Examples include hybrids of gold nanoparticles with graphene oxide (GO) or reduced GO[28], MoSe$_2$/GO[29], GO-Sb$_2$Se$_3$[30], SWCNT-VSe$_2$[31], VS$_2$-NiS$_2$[32], and perovskites[33]. However, many of these donor-acceptor complexes primarily exhibit nonlinear optical responses in the nanosecond pulse domain owing to the relatively slow charge transfer kinetics. Therefore, a significant challenge lies in identifying suitable donor-acceptor materials that can achieve superior nonlinear optical responses operating in the femtosecond pulse regime, enabled by ultrafast charge transfer processes. Finding donor-acceptor materials that exhibit improved nonlinear optical responses within femtosecond pulse regimes due to ultrafast charge transfer represents a real challenge. Overcoming this challenge is crucial to unlock the full potential of nonlinear optics and develop materials with enhanced nonlinear optical properties for ultrafast applications.

Here, we have successfully synthesized a strongly coupled hybrid material composed of VSe$_2$ hexagons and rGO sheets, which demonstrates ultrafast charge transfer from VSe$_2$ to rGO. The VSe$_2$ hexagons were chosen as the donor component of the hybrid due to their semimetal characteristics[34], high carrier mobility[35], and excellent third-order optical response[36]. Conversely, rGO sheet was selected as the acceptor material for its ease of processing and large surface area[37]. Strikingly, in the VSe$_2$-rGO hybrid, we report many-fold enhancement in the magnitude of both the imaginary and real components of the third-order nonlinear susceptibility compared to their pristine systems due to ultrafast charge transfer. For instance, we show a large ultrafast excited state absorption (ESA) coefficient ($\beta$) of (120±14) ×10$^{-3}$ cm GW$^{-1}$ and the nonlinear refractive index (50±4) ×10$^{-6}$ cm$^2$GW$^{-1}$ which are at least an order of magnitude higher than rGO. In contrast, VSe$_2$ shows only saturable absorption (SA) with negative nonlinearity (self-defocussing). The photoinduced charge transfer was experimentally verified through steady-state and time-resolved PL measurements. Furthermore, our first-principle calculations using density functional theory and Bader charge analysis confirmed the transfer of electrons from VSe$_2$ to rGO. An additional benefit of the VSe$_2$-rGO hybrid is its lower figure of merit (FOM), defined as $n_2/\beta\lambda$; a lower FOM implies that a substantial nonlinear phase shift may be accomplished in smaller length scales, making it particularly well-suited for optical limiting applications[38]. We have designed a liquid cell-based ultrafast optical limiter device using VSe$_2$-rGO hybrid, with exceptional parameters: low onset threshold (2.5 mJ cm$^{-2}$) and low differential transmittance (0.42) in femtosecond pulse domain. Our liquid-based optical limiter surpasses other benchmark devices with its superior performance parameters. Notably, it boasts self-healing capabilities, enabling high dynamic ranges while mitigating damage to cell windows.

## EXPERIMENTAL SECTION

**Powder X-ray Diffraction.** To analyze the X-ray diffraction (XRD) patterns of VSe$_2$ hexagon and VSe$_2$-rGO hybrid, we employed a Rigaku Ultima IV powder X-ray diffractometer (PXRD) equipped with a Ni-filter for Cu K$\alpha$ radiation (1.54 Å).

**Raman Spectroscopy.** The Raman spectra of the VSe$_2$-rGO hybrid and pristine samples were obtained using a Horiba LabRAM high-resolution spectrometer with a wavelength of 632.8 nm from a He-Ne laser. The spectrometer provides a resolution of approximately 1 cm$^{-1}$, allowing for detailed analysis of the Raman scattering signals.

**Z-scan measurement.** For our experimental setup, we utilized a Z-scan configuration that incorporated a Ti: Sapphire Regenerative Amplifier System with a Spectra-Physics Mai Tai Ti:

Sapphire oscillator as the seed source. To generate the excitation wavelength of 534 nm, we employed an optical parametric amplifier (TOPAS). The femtosecond laser pulses were focused onto the sample using a 30 cm lens, while the sample itself was moved along the Z-direction of the beam using a motorized translation stage. In our experimental setup, we measured the Rayleigh length ($Z_0$) to be 3.75 mm, indicating the region over which the beam remains well-focused. The beam waist ($W_0$) was determined to be 30.5 μm. To conduct the Z-scan measurements, the sample was dispersed in distilled water, which exhibited a linear transmittance of 65%.

**Computational Detail.** The calculations for geometrical optimization, optical and electronic properties were carried out by using the Vienna ab initio simulation package VASP[39–42] with the implementation of Density Functional Theory (DFT) method. The pseudopotentials used were of the projector augmented wave (PAW) type[43] with the combination of generalized gradient approximation approximation (GGA)[44] as exchange-correlation functional. To ensure the accuracy of our calculations, we set the cut-off energy for the plane-wave basis to 520 eV. For geometry optimizations the conjugated gradient method with convergence threshold of 0.01 eV/Å in force and $10^{-5}$ eV in energy were considered. The Brillouin-zone integration was sampled by the 5×5 ×1 Monk horst pack[45] grid for geometry optimization, whereas for optical and electronic properties, it was raised to 7×7 ×1. To consider the weak Van der Waals interactions between the layers of hybrid Grimme's DFT-D2[46] scheme was used.

## Result and Discussion

A strongly coupled $VSe_2$-rGO hybrid was synthesized by a facile hydrothermal method, as reported in detail in section S1 within the Supporting Information. Figure 1a presents the scanning electron microscope (SEM) image of rGO, revealing thin sheets that are randomly aggregated, with well-defined wrinkled surfaces, edges, and folding. The SEM image of pristine VSe$_2$ in Figure 1b displays the hexagonal sheet-like morphology, which is similar to previous report[47]. Figure 1c depicts the SEM image of the hybrid, clearly demonstrating the VSe$_2$ hexagons are grown onto the surface of rGO sheets which shows interconnectedness between VSe$_2$ and rGO. This interconnection enhances their interactions and strong electronic coupling. Figure S1 within the Supporting Information displays the XRD pattern of VSe$_2$ and hybrid, which precisely corresponds to the JCPDS card No.89-1641. The XRD pattern of both the samples can be shown to be identical, with no extra carbon signal in the VSe$_2$-rGO hybrid. Due to the extremely crystalline form of VSe$_2$ and the larger interplanar spacing of rGO, there are no diffraction peaks for the carbonic components in the XRD pattern of hybrid.

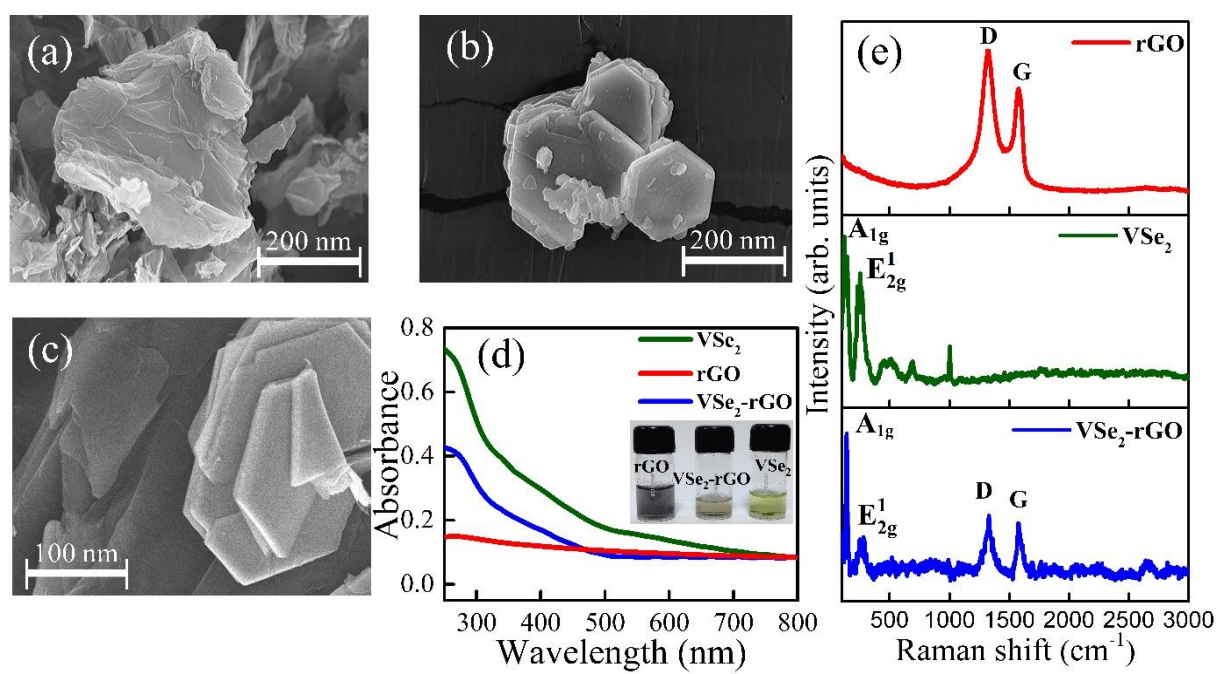

**Figure 1** Shows the SEM images of (a) rGO, (b) VSe$_2$, and (c) VSe$_2$-rGO, which display a visual comparison of their morphological characteristics. (d) Presents the ground state optical absorption spectrum of rGO (red line), VSe$_2$ (green line), and VSe$_2$-rGO hybrid (blue line). (e) Raman spectrum of rGO (upper panel), VSe$_2$ (middle panel) and VSe$_2$-rGO (lower panel).

The absorption spectra of pristine VSe$_2$, rGO and VSe$_2$-rGO hybrid were displayed in Figure 1d. The absorption onset of VSe$_2$ was found to be at 502 nm, which is quite different from the characteristic absorption feature of bulk VSe$_2$, a semimetal with absence of absorption peak within the Ultraviolet -visible region[34]. The absorption spectrum of rGO shows a characteristic absorption peak at 267 nm, ascribed to the π - π* orbital transition of C-C bonds[48,49]. Interestingly, the absorption spectrum of VSe$_2$-rGO hybrid could be understood as a superposition of signals from each pristine samples, with an overall signal that was redshifted compared to VSe$_2$, indicating strong coupling between VSe$_2$ and rGO. Figure 1e (upper panel) presents the Raman spectrum of rGO, which has two prominent modes; the disorder mode (D mode) caused by the graphite edges and defects at 1320 cm$^{-1}$ and the G mode the in-phase vibration of the graphite lattice at 1576 cm$^{-1}$. The G mode of rGO exhibits a blue-shift in comparison to the graphite mode, which is typically observed at (1587 cm$^{-1}$)[50]. The Raman spectrum of VSe$_2$ presented in Figure 1e (middle panel) possesses the characteristic A$_{1g}$ mode (Se atoms' out-of-plane vibrations) at 145 cm$^{-1}$ and the E$^1_{2g}$ mode (Atoms of V and Se vibrating in-plane, out-of-phase) at 257 cm$^{-1}$, which coincide with the earlier reports quite well[51]. Figure 1e (lower panel) displays the Raman spectrum of the VSe$_2$-rGO hybrid, pronounced as a superposition of D and G modes associated with rGO and A$_{1g}$ and E$^1_{2g}$ modes associated with VSe$_2$. However, the A$_{1g}$ mode of the hybrid is blue-shifted to 142 cm$^{-1}$, and the E$^1_{2g}$ mode is redshifted to 275 cm$^{-1}$ compared to pristine VSe$_2$. In the hybrid material, the D band of rGO experiences a redshift and appears at 1329 cm$^{-1}$. This shift indicates a strong interlayer coupling between VSe$_2$ and rGO. Interestingly, the position of the G mode of rGO remains unaffected in the hybrid.

We employed the open and closed aperture Z-scan method to investigate the third-order nonlinear optical response, which measures the overall transmittance of the sample at various positions relative to the focal point[3,31]. Further details can be found in section S3 within the Supporting Information. We dispersed the samples in water, which exhibited a transmittance of 65% for our Z-scan experiments. Figure 2a depicts the open aperture Z-scan traces of the hybrid and pristine samples under 120 fs, 532 nm excitation at a moderate peak intensity of 244 GWcm$^{-2}$. The normalized transmittance of the VSe$_2$-rGO displays a strong excited state absorption (ESA), in contrast to weak ESA observed in rGO and the saturable absorption (SA) exhibited by VSe$_2$. It is noteworthy to mention that even at below bandgap energy excitation, pristine VSe$_2$ displays SA since the material has single photon absorption states due to band-structure distortion induced by Se or V atomic deformation[52] or the coexistence of semiconducting and metallic states[53] and the peak intensity is insufficient to make the excited state absorption cross-section ($\sigma_{ES}$) more than the ground state absorption cross-section ($\sigma_{GS}$). Conversely, the weak ESA observed in rGO can be attributed to the sp$^3$ hybrid states containing oxygen functional groups[28,29]. Hence, based on our observations, we assume that the unprecedented enhancement in the ESA of VSe$_2$-rGO can be ascribed to the phenomenon of charge transfer occurring between two individual components of hybrid material. For this, we have calculated the $\sigma_{ES}$ and $\sigma_{GS}$ of the VSe$_2$-rGO and rGO, presented in Table I. From there, it is evident that $\sigma_{ES}$ of the hybrid shows a four-fold enhancement compared to rGO at a moderate peak intensity of 244 GW/cm$^{-2}$. We have performed intensity-dependent Z-scan experiments to get further insight into the ESA of the VSe$_2$-rGO. Consistent with the theoretical predictions of ESA, our measurements reveal a monotonic increase in the $\sigma_{ES}$ with peak intensity, as depicted in Table I.

**Table. I** The calculated ground state absorption cross-section ($\sigma_{GS}$) and excited state absorption cross-section ($\sigma_{ES}$) at 532 nm excitation.

| Intensity (GW cm$^{-2}$) | VSe$_2$ | | rGO | | VSe$_2$/rGO | |
|---|---|---|---|---|---|---|
| | $\sigma_{GS}\times 10^{-20}$ (cm$^2$) | $\sigma_{ES}\times 10^{-20}$ (cm$^2$) | $\sigma_{GS}\times 10^{-20}$ (cm$^2$) | $\sigma_{ES}\times 10^{-20}$ (cm$^2$) | $\sigma_{GS}\times 10^{-20}$ (cm$^2$) | $\sigma_{ES}\times 10^{-20}$ (cm$^2$) |
| 195 | 1.0 ± 0.2 | 0.3 ± 0.1 | 0.3 ± 0.1 | 1.3 ± 0.2 | 1.4 ± 0.3 | 5.2 ± 0.9 |
| 244 | 1.2 ± 0.3 | 0.4 ± 0.1 | 0.5 ± 0.1 | 1.6 ± 0.4 | 1.8 ± 0.4 | 7.6 ± 1.0 |
| 304 | 1.7 ± 0.3 | 0.9 ± 0.2 | 1.0 ± 0.3 | 3.0 ± 0.6 | 2.4 ± 0.6 | 12.0 ± 2.0 |

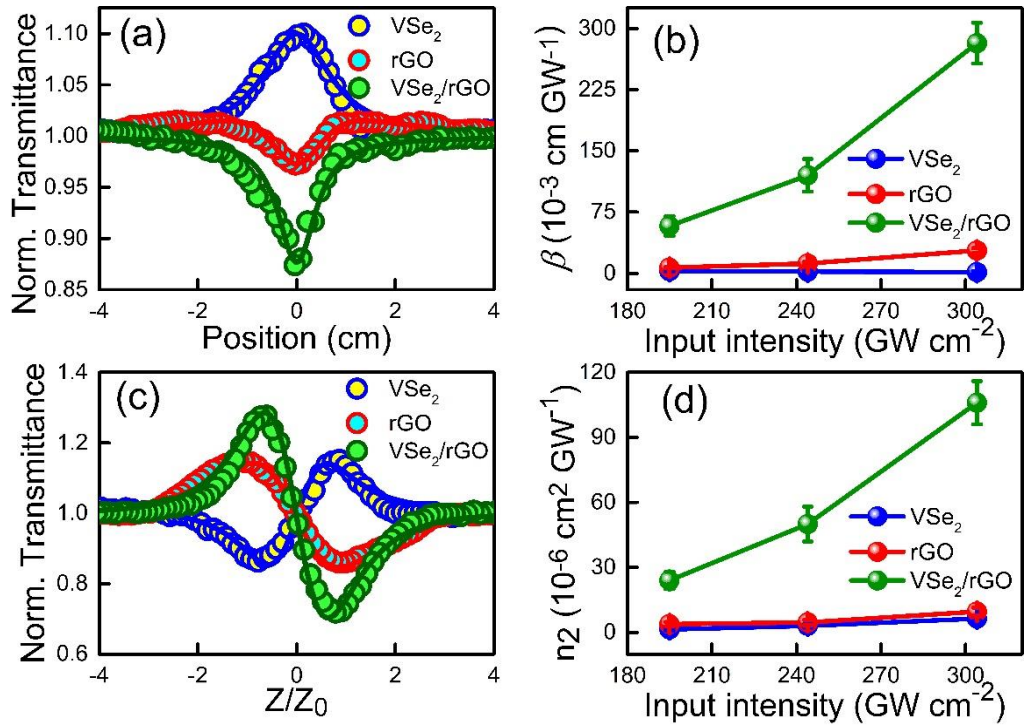

**Figure 2.** The normalized transmittance as a function of the sample position under 532 nm with a peak intensity of 244 GW cm$^{-2}$. (a) Open-aperture (c) Closed-aperture Z-scan trace of VSe$_2$, rGO and VSe$_2$-rGO hybrid. (b) Variation of $\beta$ and (d) n$_2$ with input intensity for VSe$_2$, rGO and VSe$_2$-rGO hybrid.

The Z-scan theory deliberated in the section S3 within the Supporting Information was utilized to fit the data to calculate the ESA coefficient ($\beta$) and saturation intensity ($I_s$). The calculated $\beta$ and $I_s$ values, obtained from the optimal fit to the normalized transmittance data, are presented in Table II. Strikingly, we could observe that the $\beta$ values are enhanced by as much as an order of magnitude compared to rGO. For example, $\beta = 120\pm14 \times 10^{-3}$ cm GW$^{-1}$ of VSe$_2$-rGO hybrid is ten times larger than the corresponding value for rGO. Considering the strong enhancement in the value of $\beta$ in the VSe$_2$-rGO hybrid, we propose a strong coupling and charge transfer between VSe$_2$ and rGO. Upon photoexcitation, electrons residing in the valence band of VSe$_2$ are transferred to the first excited state of rGO. Consequently, intrabank absorption from the excited state of rGO to the next higher excited state leads to strong ESA. More details regarding this process can be found in section S6 within the Supporting Information. To provide additional experimental evidence of ESA, intensity-dependent Z-scan measurements were performed. These experiments, show that $\beta$ has a monotonic increase with peak intensity, as depicted in Figure 2b.

At this juncture, it is pivotal to provide evidence supporting the notion that the discern enhancement in ESA is arising from charge transfer between the donor VSe$_2$ and the acceptor rGO. For this, we have shown that the condition for observing ESA is that the $\sigma_{ES}$ must be greater than the $\sigma_{GS}$, and if this requirement is not met, the material will display SA[28,54]. We met the criteria $\sigma_{ES} > \sigma_{GS}$ by facilitating charge transfer from the donor material to the acceptor material within the hybrid structure, clearly evident from the data presented in Table I. In this manner, we can get ESA in substances that solely exhibit SA. For example, VSe$_2$ exhibits SA that may be hybridised with electron-accepting rGO to produce enhanced ESA. These results authenticate that the enhanced nonlinear optical response of the hybrid is due to the charge transfer.

**TABLE II**. Nonlinear optical parameters: ESA Coefficient ($\beta$) and Saturation Intensity ($I_s$), nonlinear refraction (n$_2$), and figure of merit (FOM) at 532 nm excitation.

| Sample | $\beta \times 10^{-3}$ (cm/GW) | $I_s \times 10^2$ (GW/cm$^2$) | $n_2 \times 10^{-7}$ (cm$^2$/GW) | FOM |
|---|---|---|---|---|
| VSe$_2$ | -(2.2 ± 0.3) | 50 ± 5 | 3.0 ± 0.2 | 2.55 |
| rGO | 12 ± 1 | 3.5 ± 0.8 | 4.6 ± 0.3 | 0.72 |
| VSe$_2$-rGO | 120 ± 14 | 1.8 ± 0.2 | 45 ± 4 | 0.70 |

We conducted closed aperture Z-scan measurements to investigate the magnitude and sign of the third-order nonlinear refractive index (n$_2$). Figure 2c presents the nonlinear refraction profiles of VSe$_2$, rGO, and the hybrid under 532 nm excitation at a moderate peak intensity of 244 GW cm$^{-2}$. For pristine VSe$_2$, we observed a pre-focal valley and a post-focal peak, indicating a positive refractive nonlinearity and a self-focusing nature of the material. The measured $n_2$ value for $VSe_2$ is $(3.0\pm0.2)\times10^{-6}$ $cm^2$ $GW^{-1}$. This positive $n_2$ value can be attributed to bound carriers generated by nonresonant excitation[55,56]. Strikingly, an intriguing finding was observed for pristine rGO and $VSe_2$-rGO, where the Z-scan trace exhibited a peak-valley structure, indicating a negative refractive nonlinearity and a self-defocusing nature of the materials. This phenomenon arises due to the presence of free carriers and delocalized π electrons within the conduction band of the $sp^2$ domains. These components play a significant role in the nonlinear refraction and self-defocusing characteristics exhibited by rGO and $VSe_2$-rGO hybrid[31,57]. Further, the increased difference between the valley-to-peak in the Z-scan trace indicates higher $n_2$ values of $VSe_2$-rGO. We estimated the $n_2$ values by fitting the closed aperture Z-scan traces with the theoretical equations described within the Supporting Information. The measured $n_2$ values are summarized in Table I, indicating that the $VSe_2$-rGO hybrid exhibits a higher $n_2$ compared to pristine $VSe_2$ and rGO. For example, at a peak intensity of 244 $GWcm^{-2}$, the $n_2$ value of the $VSe_2$-rGO is $(45 \pm 4) \times 10^{-6}$ $cm^2$ $GW^{-1}$ is an order of magnitude larger than rGO $(4.6 \pm 0.3) \times10^{-6}$ and $(45 \pm 4) \times 10^{-6}$ $cm^2$ $GW^{-1}$. Additionally, we observed an intensity-dependent variation of $n_2$, which displayed a monotonic increase with peak intensity shown in Figure 2d. Moreover, a figure of merit (FOM) defined as $n_2/\lambda\beta$, is a parameter that embodies the performance of a device. Here, the low figure of merit (FOM) of $VSe_2$-rGO (see Table II) suggests that the hybrid can achieve a significant nonlinear phase shift even in smaller length scales, making it highly suitable for optical limiting.

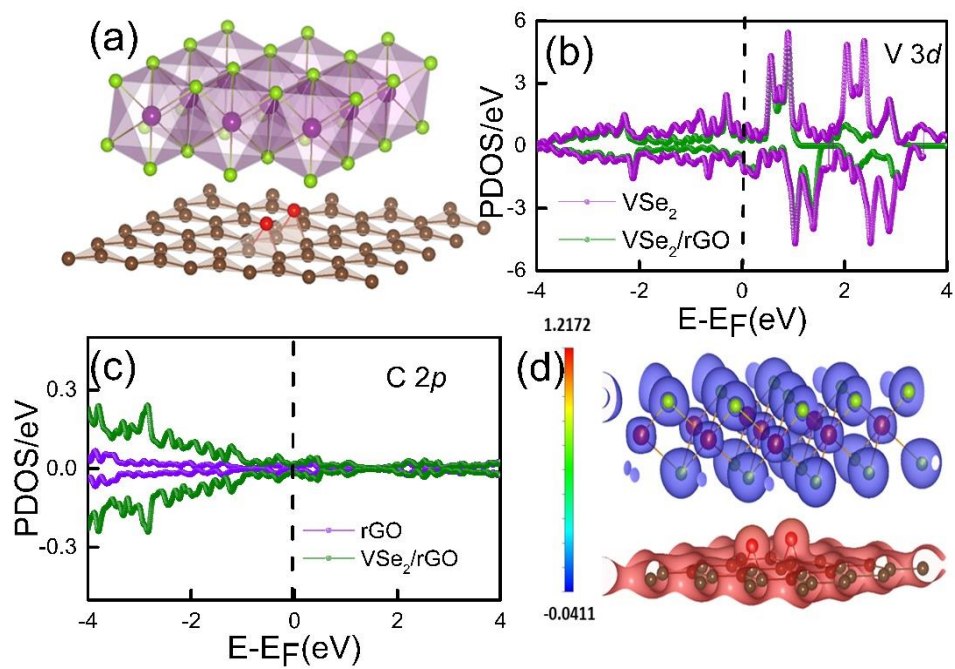

**Figure 3.** (a) Optimized structure of hybrid of VSe$_2$ (001) with rGO. (b) PDOS plot for V 3d orbital of VSe$_2$ before and after hybridization with rGO and (c) PDOS plot for C 2p orbital of rGO before and after hybridization with VSe$_2$; Purple, Green, Brown, and Red indicate V, Se, C and O atoms respectively. (d) The charge density difference between VSe$_2$-rGO and rGO, with an isovalue of 0.08e. In the plot, the regions where VSe$_2$ loses charge are depicted in blue, while the regions where rGO gains charge are represented in red.

To gain further understanding into the charge transfer phenomenon in the VSe$_2$-rGO hybrid, we performed computational calculations using Density Functional Theory (DFT) employed in the Vienna Ab initio Simulation Package (VASP)[39–42]. These calculations are detailed in Section S7 within the Supporting Information. The DFT-optimized structures of VSe$_2$ and rGO with (001) surface is displayed in Figure 3a, while the TDOS for VSe$_2$-rGO is determined in Figure S8 within the Supporting Information. The TDOS of the hybrid shows an enrichment of states near the Fermi energy level in comparison to pristine VSe$_2$ owing to the increased surface area subsequent from the introduction of rGO, which offers more active sites for charge transfer from nanoparticles and causes to the observed modifications in the electronic structure of the hybrid material. To account for orbital interactions and charge transfer phenomena, we plotted the partial density of states (PDOS) for the V 3d and C 2p orbitals of $VSe_2$ and rGO. The PDOS plots, displayed in Figure 3b and c, provide insights into the distribution of electronic states within these specific orbitals before and after the hybridization process. The V 3d orbital of $VSe_2$ exhibited an overall reduction of states at and below the Fermi level after hybridization with rGO, as seen in Figure 3b. In contrast, the C 2p orbital of rGO showed continuous enhancement of states below the Fermi level, as seen in Figure 3c. Therefore, we can conclude that there is a charge transfer from $VSe_2$ to rGO, which supports our experimental findings. To corroborate our findings, we estimated the quantitative Bader charge partitioning to analyze the charge transfer qualitatively[58]. This analysis revealed a net charge transfer of 0.10e from $VSe_2$ to rGO, with the V 3d orbital exhibiting the maximum loss of charge. This is in agreement with the qualitative PDOS analysis discussed earlier. To visualize and examine the spatial distribution of charge transfer, we generated a plot illustrating the charge density difference between $VSe_2$-rGO and rGO shown in Figure 3d. Here, the charge gained by rGO is represented in red, while the charge lost by $VSe_2$ is represented in blue.

To uncover further the charge transfer mechanism, we have performed PL measurements on the hybrid and pristine $VSe_2$, as depicted in Figure 4a. The results revealed that $VSe_2$ was highly photoluminescent, exhibiting a broad peak at the position of 456 nm. Remarkably, the PL intensity was significantly quenched by a factor of 3.4 and redshifted by 11 nm in the hybrid in comparison to $VSe_2$, demonstrating an effective charge transfer. The covalent bonding in the hybrid modifies the electronic properties of rGO and $VSe_2$, leading to quenching and red shifting of the PL peak[59]. Moreover, we conducted time-resolved photoluminescence (PL) measurements utilizing a 298-nm laser excitation and 456-nm emission, as depicted in Figure 4b. The decay curves for both the hybrid material and pristine $VSe_2$ were accurately fitted using exponential functions, and the corresponding fitting parameters can be found in Table S4 within the Supporting Information. The average PL lifetime of the pristine VSe$_2$ was found to be (0.91 ± 0.05) ns, which drastically reduces to (0.31 ± 0.02) ns in VSe$_2$-rGO, indicating efficient charge transfer.

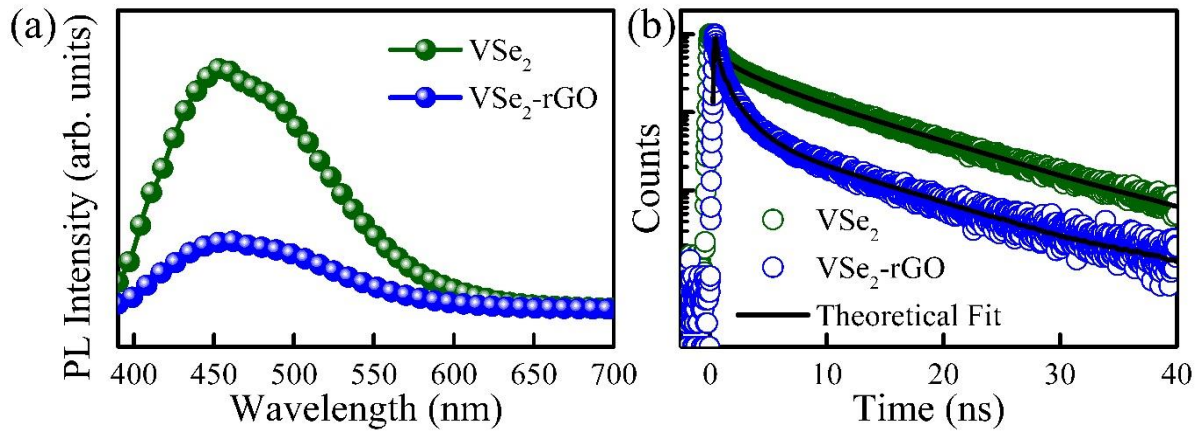

**Figure 4.** (a) PL spectra of VSe$_2$ and VSe$_2$-rGO hybrid. The PL intensity of the hybrid material exhibits a significant quenching, accompanied by a redshift of 11 nm in comparison to pristine VSe$_2$, demonstrating an effective charge transfer. (b) The decay lifetime of control VSe$_2$ and the VSe$_2$-rGO hybrid.

The remarkable enhancement in the ESA and the nonlinear refraction of hybrid suggests that it has the potential to be used as an ideal optical limiter. Henceforth, we have designed a liquid cell-based optical limiter device, utilizing an aqueous VSe$_2$-rGO as the medium with linear transmittance of 65% depicted in Figure 5a. We evaluated the performance of our device using a Z-scan setup, which allowed us to assess the optical limiting capabilities. The effectiveness of our device was determined by analyzing two key parameters: the onset threshold ($F_{ON}$), which represents the input intensity where the transmittance deviates from linearity. and the differential transmittance ($T'_{diff}$), calculated as the derivative of the output intensity ($dI_{Out}$) with respect to the input intensity ($dI_{In}$), at high intensity[28,31]. A Low $F_{ON}$ and $T'_{diff}$ were desirable characteristics for an ideal optical limiter. Figure 5b shows the normalized transmittance as a function of input intensity, and output versus input fluences are plotted in Figure 5c. From these plots, our measurements show $F_{ON}$ of 2.5 mJ cm$^{-2}$ and T'$_{diff}$ of 0.42. In Table III, we compared the performance of our device with that of existing optical limiters operating in the nearly same wavelength range. This table shows that measured device parameters are better than several benchmark optical limiters, demonstrating that our device is an excellent optical limiter. Further, our liquid-based optical limiter possesses a significant advantage in that it has the ability to self-heal, enabling it to handle high dynamic ranges without constraint, except for potential damage to the cell windows.

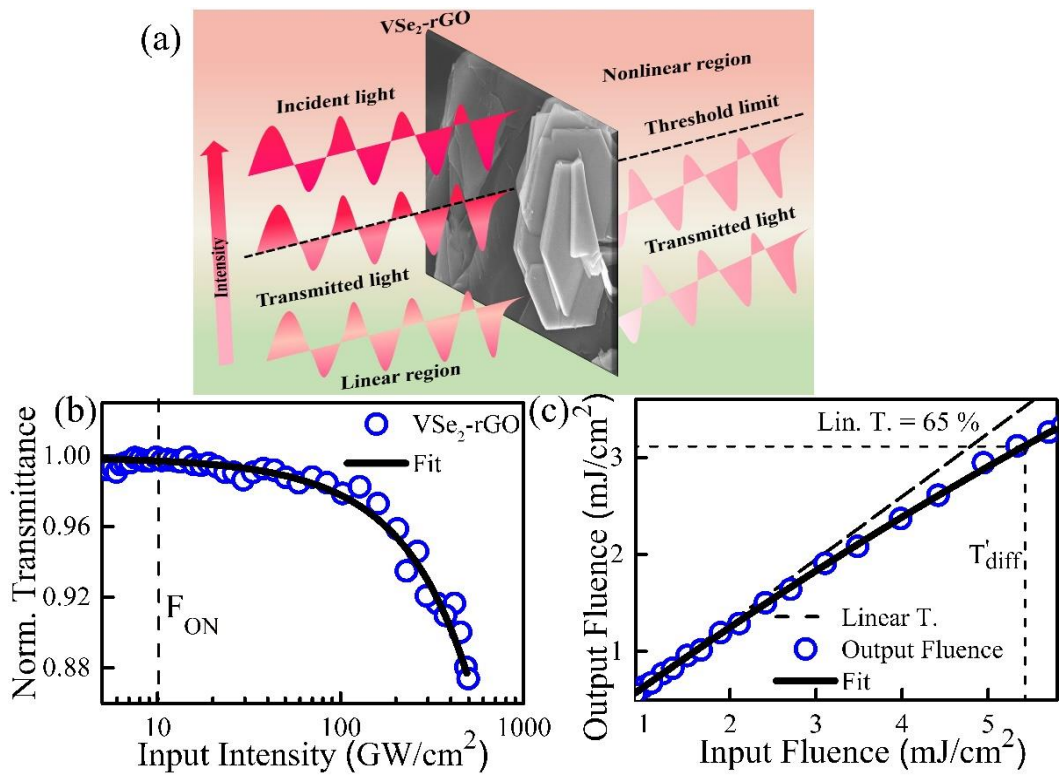

**Figure 5.** (a) Schematic of optical-limiting device contains VSe$_2$-rGO aqueous solution as optical limiting medium. The limiting medium distilled water has the linear transmittance of 65% (red zone, dotted line). Our device operates in such a way that it permits the passage of light when lower-intensity beams are within the linear zone (green region). However, when a powerful beam exceeds the threshold limit (red zone, dotted line), the device functions as an optical limiter, attenuating its intensity and preventing it from causing damage or undesirable effects. (b) Normalized transmittance as a function of input intensity with 120 fs, 534 nm excitation. (c) Output fluence vs input fluence to determine T'$_{diff}$, in figure the linear transmittance as dashed line and theoretical fitting as solid lines, respectively.

**TABLE III**. Nonlinear optical limiting parameters: the onset threshold ($F_{ON}$) and the differential transmittance ($T'_{diff}$) in comparison to several benchmark optical limiters.

| Systems | Wavelength (nm) | Linear T. | $F_{ON}$ (mJ/cm$^2$) | $T'_{df}$ | References |
|---|---|---|---|---|---|
| VSe$_2$-rGO | 534 | 65 | 2.5 | 0.42 | Present work |
| Au NP-rGO | 532 | 70 | 50.0 | 0.3 | Ref[29] |
| C60 | 532 | 63 | 65.0 | - | Ref[60] |
| Graphene thin film | 532 | 73 | 10.0 | 0.1 | Ref[61] |
| Graphene nanoribbon | 532 | 70 | 100 | - | Ref[62] |
| CDS nanoparticles | 532 | 70 | 300 | 0.56 | Ref[63] |

## Summary

In summary, we demonstrated an ultrafast third-order nonlinear absorptive and refractive optical response in the VSe$_2$-rGO hybrid. We found that the synergistic charge transfer between the VSe$_2$ and rGO of the hybrid resulted in a manifold enhancement in the ESA coefficient and nonlinear refractive index exhibiting a self-defocusing nature. Our computations, through DFT and Bader charge analysis confirms the charge transfer from VSe$_2$ to rGO. We also designed an optical-limiting device that operates in visible range over femtosecond pulses, and we found that it had important device parameters on par with several other benchmark optical limiters. Our research opens us new possibilities for adjusting and managing the ultrafast third-order nonlinear optical response and the development of future optoelectronic devices based on hybrid materials.

## Acknowledgements

The authors gratefully acknowledge the Science and Engineering Research Board (project no. EMR/2016/002520 and CRG/2019/002808), DAE BRNS (sanction no. 37(3)/14/26/2016-BRNS/37245), and FIST Project for Department of Physics. BC acknowledges Dr. Nandini Garg, Dr. T. Sakutnala, Dr. S.M. Yusuf, and Dr. A.K. Mohanty for support and encouragement. CSR gratefully acknowledge financial assistance from the SERB Core Research Grant (Grant No. CRG/2022/000897), Department of Science and Technology (DST/NM/NT/2019/205(G)), and Minor Research Project Grant, Jain University (JU/MRP/CNMS/29/2023).